# The Interaction of Matter and Radiation:
# The Physics of C.V. Raman, S.N. Bose and M.N. Saha
# Part 2: Physics Highlights

Arnab Rai Choudhuri

Department of Physics

Indian Institute of Science

Bengaluru – 560012

e-mail: arnab@iisc.ac.in

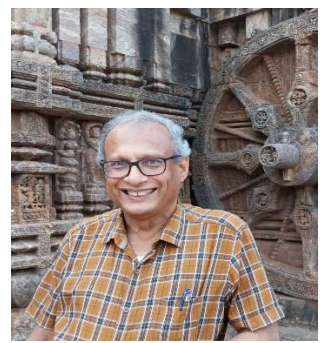

The author is a theoretical astrophysicist working at the Department of Physics, Indian Institute of Science. He was involved in developing the currently accepted theoretical model of the 11-year sunspot cycle. He also has interest in the history and philosophy of science.

**Summary.** Three extraordinary physics discoveries were made from colonial India, which did not have any previous tradition of research in modern physics: Saha ionization equation (1920), Bose statistics (1924), Raman effect (1928). All the three discoverers were founding faculty members of the new small physics department of Calcutta University, which started functioning from 1916. These discoveries were all in the general topic of interaction between matter and radiation. In Part 1, we have described the social and the intellectual environment in which these discoveries were made. Now, in Part 2, we shall first give a background of the revolutionary developments taking place in physics at that time. Then we shall provide a non-technical account of the three discoveries and point out the kind of impact these discoveries made in the subsequent development of physics.

**Keywords:** Culcutta University – Raman effect – Bose(-Einstein) statistics – Saha ionization equation

**Introductory Remarks**

Chadrasekhara Ventata Raman (1888 – 1970), Satyendranath Bose (1894 – 1974) and Meghnad Saha (1893 – 1956), working under very adverse conditions in colonial India, reached extraordinary heights of scientific creativity in physics approximately a century before the present time. The aim of the present article in two parts is to provide a non-technical account of their three famous discoveries – Saha ionization equation (1920), Bose statistics (1924), Raman effect (1928) – suitable for readers without a technical background in physics.

In Part 1, I have already given a historical account of the conditions of the society in which these extraordinary physics discoveries were made. It was also mentioned towards the end of Part 1 that all these discoveries were in the general area of the interaction between matter and radiation – the central theme in the physics of that era which led to the quantum revolution. This makes it scientifically meaningful to discuss these three discoveries together within one article. I shall now say a few words about matter, then about radiation and finally about their interaction, before I discuss the discoveries of Saha, Bose and Raman.

**Scientific Background: Matter, Radiation and their Interaction**

*Crash course on matter*

Many ancient philosophers speculated that matter is made up of atoms. However, only during the first decade of the 19$^{th}$ century, Dalton [1] formulated modern atomic theory to explain certain regularities in chemical reactions. Atoms were regarded as indivisible until J.J. Thomson [2] discovered the electron, a sub-atomic particle, in 1897. The question of how electrons are organized inside an atom was finally answered by Rutherford [3] in 1911 through an ingenious experiment. He argued that an atom consists of a positively charged nucleus with negatively charged electrons going around. In other words, the interior of an atom is like a miniature version of our solar system. However, Bohr [4] pointed out an important difference in 1913. A planet can be at any distance from the Sun without violating any law of physics. To explain various kinds of experimental data, the Bohr model of 1913 suggested that electrons can occupy only certain specific orbits around the nucleus – leading to certain discrete values of the energy of the atom.

*Crash course on radiation*

The most common example of radiation is light, through which we see objects around us. The nature of light has been debated by philosophers and scientists for a long time. Newton [5] suggested that light consists of corpuscles. However, in 1803 Young [6] discovered what is called interference of light, establishing the wave theory of light. A burning question about light was: what is it wave of? Maxwell [7] showed in 1866 that light is an electromagnetic wave, consisting of rapidly oscillating electric and magnetic fields. A wave means that something would be oscillating at a place with certain frequency. For yellow light, the frequency is $v$ ~5 x $10^{14}$ per second. Radio waves have lower frequency of about $v$ ~$10^8$ per second, whereas X-rays have higher frequency of about $v$ ~$10^{18}$ per second.

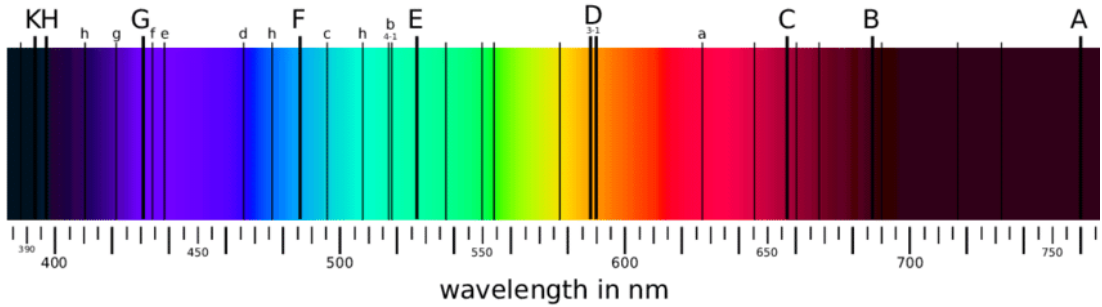

Figure 1. The spectrum of sunlight with the dark Fraunhofer lines. Source: Wikipedia commons.

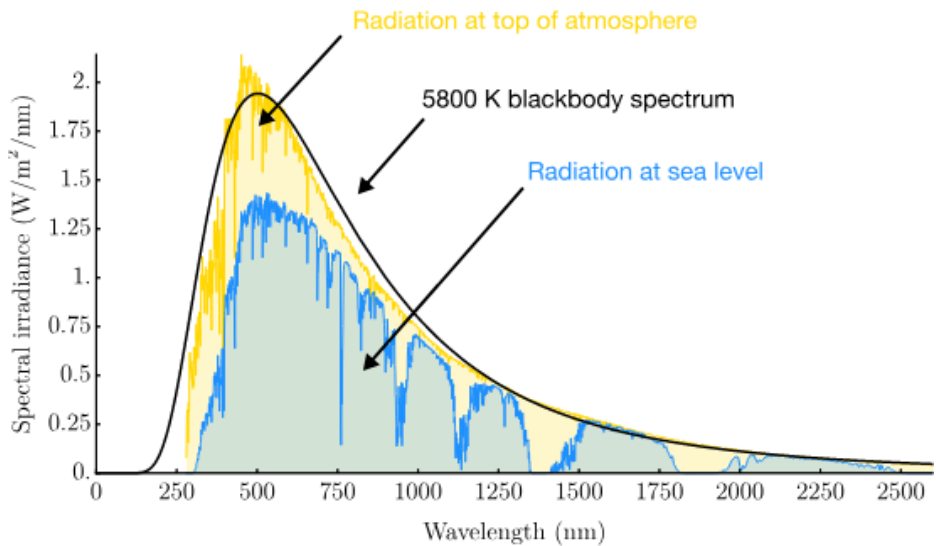

Figure 2. A graphical representation of the spectrum of sunlight, in which the energy associated with a wavelength is plotted against the wavelength. The spectrum at the sea level (shown in blue) is somewhat different from the spectrum of pristine solar radiation at the top of the atmosphere, since some radiation gets absorbed in the atmosphere. The blackbody spectrum for temperature 5800 K shown by the black curve fits the spectrum at the top of the atmosphere, except for the spectral lines. Source: Wikipedia commons.

### *Interaction between them*

We know from everyday life that matter and radiation interact with each other. Hot matter like a heated piece of iron emits radiation. We also know that, when sunlight is passed through a prism, it breaks into a spectrum of different colours. Fraunhofer [8] discovered dark lines in the solar spectrum in 1815. These dark lines can be seen in the spectrum of sunlight shown in Figure 1. Bunsen & Kirchhof [9] realized in 1861 that these dark spectral lines are signatures of different chemical elements. While non-scientists may like to see a colourful picture of a spectrum like Figure 1, scientists often prefer to show the spectrum in the form of a more boring graph, in which we plot the energy associated with different wavelengths or frequencies. Such a graph of sunlight spectrum is shown in Figure 2. If you see the spectrum at the top of the atmosphere, you notice that it has a smooth part which is approximately like what is called a blackbody spectrum. I am going to explain this term 'blackbody spectrum' in the next paragraph. Superimposed on this smooth spectrum, you see several sharp dips corresponding to the spectral lines. I may mention that S.N. Bose's work centred on explaining the smooth blackbody spectrum, whereas M.N. Saha's work centred on explaining the spectral lines.

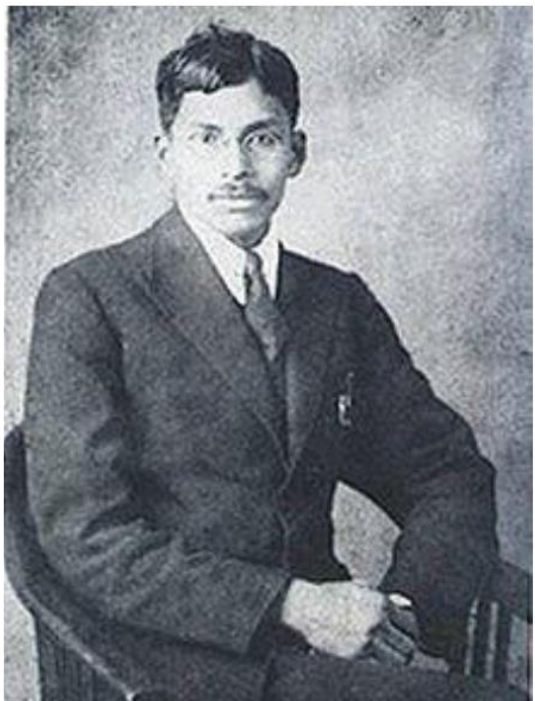 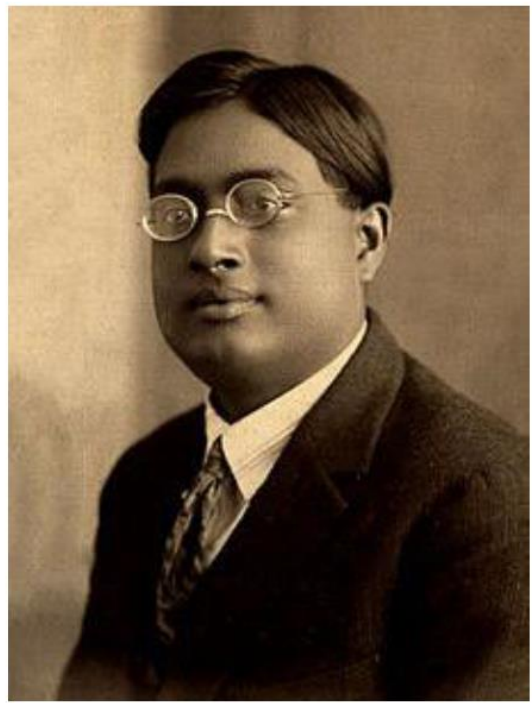

Figure 3. M.N. Saha and S.N. Bose photographed around the time when their famous works were done.

Let me now tell you what blackbody radiation or blackbody spectrum is. Suppose we make hollow boxes with different metals – iron, copper, aluminium – and keep all of them at the same temperature $T$. We then make a small hole on the side of a metal box so that a little bit of radiation from the interior comes out and we take a spectrum of this radiation by making it pass through a prism. We will find that the spectra from all the boxes made with different metals will be identical. Basically, inside the hollow boxes, we have radiation in equilibrium with matter. This radiation in equilibrium with matter, which we call blackbody radiation, does not depend on what the box is made of and is always the same for a particular temperature $T$. Since the radiation from the solar surface was approximately in equilibrium with the matter at the surface before coming out, it has an approximate blackbody spectrum. However, atoms in the surface layer of the Sun absorb radiation at certain specific wavelengths, giving rise to the spectral lines.

The quantum theory was first invented by Planck [10] in 1900 to explain why the blackbody spectrum looks as it is. Planck showed that the blackbody spectrum can be explained by postulating that matter absorbs or emits radiation of frequency $v$ only in quanta having energy $hv$. Initially, it was thought that we need to bother about quanta only when we consider absorption or emission of radiation by matter. However, Einstein [11] realized in 1905 that some experimental results can be explained only if radiation propagates also in the form of quanta called photons. Finally, Bohr [4] gave an elegant explanation of spectral lines by using his model in which atoms have discrete energy levels, as already mentioned. An atom in a state with energy $E_{\text{lower}}$ can make a transition to a state with energy $E_{\text{higher}}$ by absorbing a photon with frequency $v$ given by

$$E_{\text{higher}} - E_{\text{lower}} = hv$$

In other words, the atom would absorb only photons of a particular frequency, giving rise to a spectral line. The quantum theory eventually led to the more sophisticated quantum mechanics developed from 1925. But we do not have to get into a discussion of quantum mechanics for an understanding of the works of Saha, Bose and Raman, to which we turn now.

## Saha Ionization Equation

We begin our discussion of the Saha ionization equation by showing in Figure 4 how the equation looked in the first paper of Saha on this subject [12]. After Rutherford [3] established the nuclear model of the atom in 1911, it became clear that under certain circumstances an electron may come out of the atom. This would make the remainder part of the atom, called the ion, positively charged. This process is known as ionization. As the temperature of a gas is raised, more and more atoms get ionized. The Saha equation gives the fraction $x$ of atoms which would be ionized at a certain temperature $T$ and pressure $P$. For some given $T$ and $P$, we can calculate the ionization fraction $x$ from this equation, provided we know $U$ which denotes the energy to be supplied to an atom for it to be ionized. Saha figured out how $U$ can be estimated for several atoms from the data of some atomic physics experiments. It can be shown from the Saha equation that the ionization fraction $x$ increases with temperature and eventually the whole gas gets ionized at a sufficiently high temperature. Such an ionized gas is called a plasma. The Saha equation thus describes the process by which an ordinary gas gets converted into a plasma on increasing the temperature and is a fundamental equation in plasma physics.

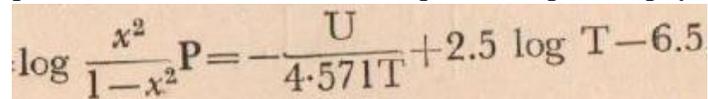

$$\log \frac{x^2}{1-x^2} P = -\frac{U}{4 \cdot 571 T} + 2.5 \log T - 6.5$$

Figure 4. The famous Saha ionization equation as written down in the first paper of Saha [12] on this subject.

Saha applied this equation to solve some of the frontier problems of astrophysics of his day. Even looking at stars through the naked eye, we become aware that different stars have different colours: some appear reddish, while others may appear bluish. Astronomers discovered that stars of different colours have different sets of spectral lines. Since spectral lines indicate the compositions of stars, astronomers were confronted with a peculiar question: do stars of different colours have different composition? Saha realized that stars of different colours have different surface temperatures, which will cause different levels of ionization. Even if these stars have the same composition, the differing levels of ionization may give rise to differing sets of spectral lines. Saha's work suddenly clarified all the mysteries of stellar spectra. Suddenly everything fell in place.

Saha is considered one of the founding fathers of two important branches of physics: plasma physics and astrophysics. Rosseland wrote in his well-known 1936 textbook *Theoretical Astrophysics* [13]:

> The impetus given to astrophysics by Saha's work can scarcely be over-estimated, as nearly all later progress in this field has been influenced by it, and much of the subsequent work has the character of refinements of Saha's ideas.

Saha was nominated for the Nobel Prize in 1930, 1937, 1939, 1940, 1951, 1955. Since Nobel Prize was usually not given for works on astrophysics at that time, Saha did not have too much chance. Still, it is interesting that Saha was one of the three shortlisted candidates in 1930 – the year in which the Nobel Prize was given to Raman [14].

**Bose(-Einstein) Statistics**

Coming to Bose now, he is the person who gave the first systematic derivation of the blackbody spectrum. It was widely recognized that Planck's famous 1900 derivation [10] used the idea of the quantum in an ad hoc manner and was not fully satisfactory. But, before Bose's work, nobody could figure out how to give a proper derivation of the blackbody spectrum. To understand Bose's approach, let us first consider molecules in a volume of air. The molecules in the air move around randomly. We do not find all the molecules moving at the same speed, but there is a distribution of speeds: some molecules moving faster than the average, whereas some moving slower. Why is there such a distribution of speeds instead of all the molecules moving with the same speed? One can show that the probability of all molecules moving with the same speed is rather low. There is a branch of physics – known as statistical mechanics – in which theoretical methods are developed for calculating such probability. Some of the greatest physicists of the late 19th century – Maxwell, Boltzmann, Gibbs – developed these theoretical methods. After Einstein's 1905 work [11] that light would consist of photons, it became possible to think of blackbody radiation as a gas of photons. Can one apply methods of statistical mechanics to this gas of photons? Since photons are quantum particles, Bose realized that one has to do the calculations differently from the way they are done for classical particles. Bose figured out the correct way of doing this calculation and succeeded in deriving the blackbody spectrum correctly. Thus, Bose simultaneously had two spectacular achievements. On the one hand, he gave the first fully consistent theoretical derivation of the blackbody spectrum. On the other hand, he figured out how to do statistical mechanics with quantum particles, leading to a new branch of physics – quantum statistical mechanics.

When Bose's small paper was rejected by *Philosophical Magazine*, he sent it to Einstein with a covering letter (shown in Figure 5), asking Einstein whether he considered the paper worth publishing in a German journal. Einstein was so amazed by this paper from a stranger that he himself translated the paper into German and got it published in *Zeitschrift fur Physik* [15]. Bose applied his statistical method only to particles like photons which have zero mass. Einstein quickly realized that the method can be generalized to include particles with non-zero mass [16]. The statistics with Einstein's generalization to particles with non-zero mass came to be known as the Bose-Einstein statistics. Einstein got a peculiar result for particles with mass, known as the Bose-Einstein condensation, which was verified experimentally much later. It was also realized by Fermi and Dirac that quantum particles like electrons would follow a somewhat different statistics, now known as the Fermi-Dirac statistics. While particles obeying the Bose-Einstein statistics are called bosons, particles obeying the Fermi-Dirac statistics are called fermions. Any elementary particle has to be either a boson or a fermion.

Figure 5. S.N. Bose's covering letter to Einstein when sending his paper to Einstein.

In the famous biography of Einstein titled *Subtle is the Lord*, Abraham Pais [17] listed four revolutionary works in old quantum theory: (i) Planck's 1900 work proposing the quantum to explain blackbody radiation [10]; (ii) Einstein's 1905 work that light consists of photons [11]; (iii) Bohr's 1913 work proposing his model of atomic structure and spectral lines [4]; (iv) Bose's 1924 work initiating quantum statistical mechanics [15]. Probably any professional physicist will agree with Pais's list of four most important works of old quantum theory. Apart from Bose's work, the other three works were awarded the Nobel Prize. Although Einstein's work on relativity may be much more famous, he got the Nobel Prize for his work on photons. In the case of Planck and Bohr, who were legendary figures in the 20[th] century physics, the papers Planck [10] and Bohr [4] listed here are their most famous papers, on which their fame principally rests. There cannot be a higher praise for a physics paper than to be classed with three other such papers. Surprisingly, Bose was nominated for the Nobel Prize after 1956 only by Indians and received the other important honour of being elected FRS (Fellow of the Royal Society of London) also only in 1958 when Dirac visiting India came to know that Bose was not an FRS and took steps to nominate him. Raman and Saha were elected FRS reasonably early – in 1924 and 1927 respectively. I may point out that several Nobel Prizes had the name Bose or boson in their citations: (i) W and Z bosons (1984); (ii) Bose-Einstein condensation (2001); (iii) Higgs boson (2013). I am not suggesting that Bose deserved any direct credit for these works. I am merely pointing out the names of these Nobel winning works.

**Raman Effect**

Coming to Raman at last, it is interesting to contrast him with Saha and Bose before discussing his science. Saha was 26 and Bose 30 at the time of their famous works. While those were not their first papers, their earlier papers were not of much consequence and they were relatively unknown in the physics community when their famous works were published. Raman, who made his famous discovery at age 40, was already established for his works on acoustics and optics, and was already FRS. Saha and Bose were theorists who dabbled in experiments in later life, whereas Raman was an experimenter who dabbled in theory in later life. Saha and Bose were local boys who spoke Bengali, the language spoken in the eastern region of India around Calcutta. But Raman came from the Madras Presidency in the south of India and spoke Tamil. A brilliant student, Raman had to join the finance department due to the lack of career opportunities in physics in India of that era and came to Calcutta, the capital of British India, in 1907 to work as a finance officer. While going to work by tram one day, Raman saw the signboard of Indian Association for Cultivation of Science (IACS) and went inside to investigate what it was. Mahendra Lal Sircar, the founder, was dead by then. Raman met his son. Raman told the junior Sircar that he had already published some papers in international journals and wanted to know if he could work in the laboratory of the Association. The junior Sircar rose from his seat, embraced Raman and said that they had been waiting for somebody like him all these years. As I already mentioned, Asutosh Mookerjee appointed Raman as Palit Professor of Physics at Calcutta University on the basis of the work he was doing at IACS.

An ambitious man, Raman was perhaps a little jittery that he had not yet discovered something truly extraordinary till about age 40. Also, he was regarded as somebody who was doing old-fashioned 19[th]-century physics like acoustics and optics, but ironically it was he who provided a definitive experimental proof of the new quantum mechanics. It is outside the scope of this article to describe the epic journey of Raman from being an old-fashioned physicist to somebody who provided the experimental proof of the newest theory in physics. I can only briefly explain his discovery. Suppose light of frequency $v_0$ passes through a liquid with molecules having energy levels $E_{higher}$ and $E_{lower}$. Theoretical calculations based on quantum mechanics suggested that a photons can absorb energy from a molecule at higher level so that the molecule goes to the lower level and the photon takes away the excess energy, which increases the frequency of the photon to $v_+$:

$hv_+ = hv_0 + (E_{higher} - E_{lower})$

A photon can also give out energy to a molecule at the lower level to come out with reduced frequency $v_-$:

$h\nu_- = h\nu_0 - (E_{\text{higher}} - E_{\text{lower}})$

Figure 7 shows Raman's own drawing of his experimental setup. He broke the bright tropical sunlight into a spectrum with the help of a prism and then made light of a single colour pass through a benzene solution. He found faint light coming out with two new colours (i.e. two new frequencies $\nu_+$ and $\nu_-$), as predicted by quantum mechanics. As may be expected, other groups were also on the trail. A Russian group headed by Mandel'shtam also discovered the effect almost at the same time, leading to a priority dispute [18].

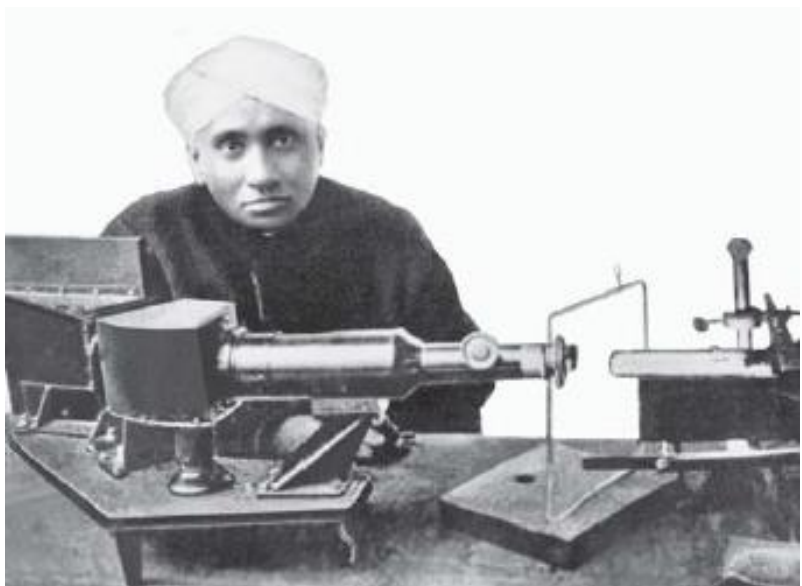

Figure 6. C.V. Raman with his spectroscope used in his discovery.

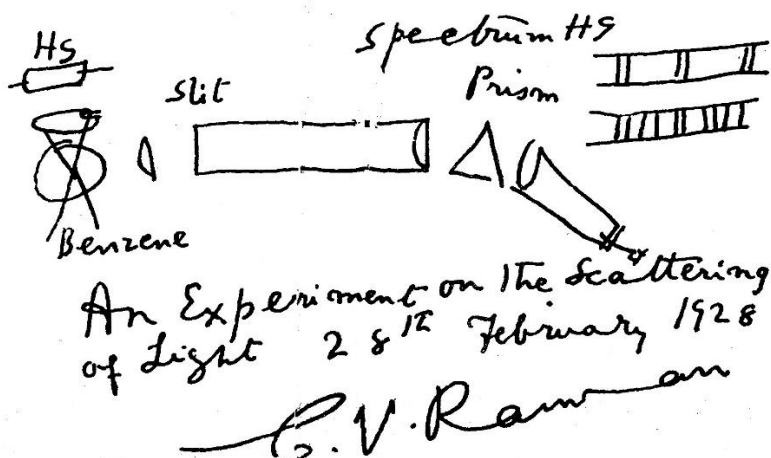

Figure 7. Raman's own sketch of his experimental setup. Courtesy: Rajinder Singh.

There has been another more serious dispute: should K.S. Krishnan, Raman's student and co-discoverer [19], get more credit for the work than what he has been given? Most of the papers in the discovery series were joint papers by Raman and Krishnan. Figure 8 shows an extract from Krishnan's diary of 9 February 1928, the day when Raman realized that they were on the threshold of a big discovery. Krishnan wrote [20]:

When Prof[essor] returned after his walk he told me that I ought to tackle big problems like that. . . Told Mr Venkateswaran about the discovery and was discussing the problem with us, in the course of which he said that the phenomenon should be called the Raman-Krishnan-Effect.

When a supervisor suggests a research problem to a student leading to a major discovery, it is always a tricky question as to who should get how much credit. I may mention that Krishnan's diary is mysteriously found in a torn condition, ending abruptly in the middle of a sentence during the entry for 28 February 1928, the recognized day of the discovery.

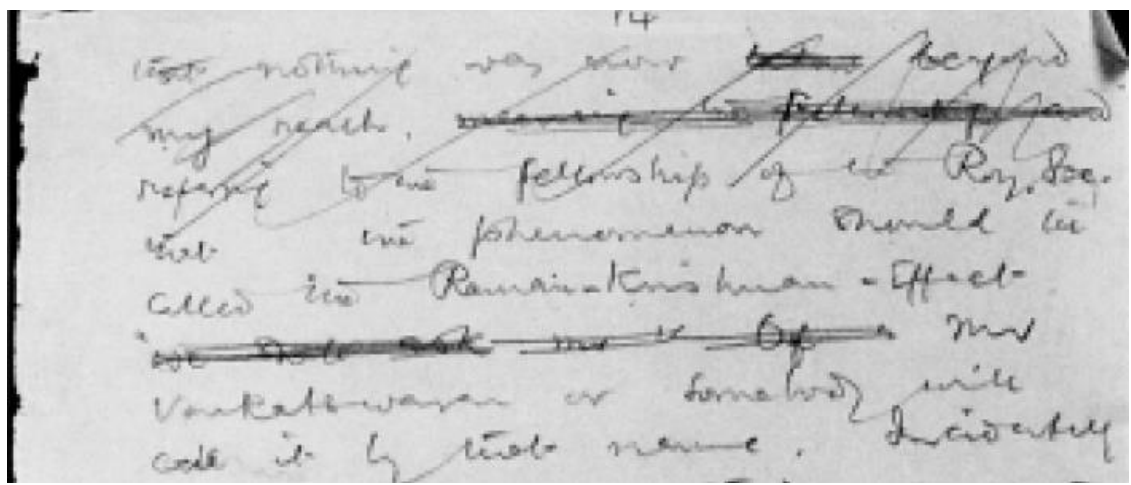

Figure 8. The portion of Krishnan's diary where Krishnan refers to Raman mentioning that the effect would be known as "the Raman-Krishnan-Effect". Courtesy: D.C.V. Mallik.

The importance of the discovery was realized immediately. In a letter dated 16 June 1928 – less than four months from the date of the discovery – Sommerfeld wrote to Raman [21]:

Your papers formed the subject of a lecture which I gave in our colloquium at Munich. . . I congratulate you most heartily on this great result. Japan, that has a for a much longer time worked in Modern Physics, has nothing to show comparable with this discovery.

Raman was so confident of winning the 1930 Nobel Prize that he booked a sea passage to Stockholm even before the announcement! In those days, nearly all the winners used to be Europeans and there would not be enough time after the announcement for somebody in a distant land to book a sea passage. Rabindranath Tagore, the first Asian to win the Nobel Prize, could not attend the Nobel ceremony. But Raman, the supreme showman, did not want to miss it!

**Concluding Remarks**

After giving an account of the intellectual atmosphere of the society in which Raman, Bose and Saha worked, I have given a non-technical description of their scientific achievements. The importance of a scientific work cannot be assessed by looking at it in isolation. A scientific work appears in a scientific tradition and its importance depends on whether it has been able to provide important answers to some important questions which arose in that scientific tradition. That is why I tried to explain the specific questions in the general area of interaction between matter and radiation which Raman, Bose and Saha addressed, and have also indicated how their works were viewed by the contemporary scientific community. In this short article, it has not been possible to give any account of the lives and careers of our protagonists. Readers are referred to [22] for an extensive bibliography of the secondary sources on these scientists.

I have pointed out that a succession of three such important works of modern physics coming out of a country without any previous tradition of physics research – and far away from the scientific centres where a scientific revolution was going on – is something unique and without parallel in the annals of the history of science. I have tried to describe the conditions of the society which provided an opportunity for a few extraordinary individuals to reach the highest peaks of scientific creativity. However, in order to have a better understanding of the mystery of such unusual creativity, one also needs to look at the mental processes of these individual scientists which led to their extraordinary achievements. It is not easy for an individual in a society without a scientific tradition to develop the appropriate mental aptitude for research [23]. Usually, in a regular scientific paper presenting a finished scientific work, the history of the struggles of the individual scientist to arrive at the final result is left out. A historian of science often has to look at a scientist's private papers to reconstruct a history of the creative process – his/her exchange of letters with other scientists, notes, reports on works in progress, official correspondence with the employer and funding agencies, sometimes even personal letters to family members and friends. That is why private papers of creative individuals are of utmost importance as inputs in an intellectual history. Many universities and societies in Europe and North America have well-funded and well-managed archives in which private papers of creative persons associated with these organizations are preserved and are made available to qualified scholars who want to study them. A scholar wishing to do research on an important European or American scientist has to approach the archives where his/her private papers are preserved.

In India, documents connected with administrative and political history – including private papers of important political leaders like Gandhi and Nehru – have been preserved in various national and state archives. Such documents are indisputable resources for historical research. Unfortunately, no such systematic efforts have been made to preserve the private papers of even the most important Indian scientists and documents connected with the establishments of important scientific institutions. India never developed a strong tradition in the history of science and there has never been a national initiative to preserve such documents essential for the study of the scientific growth of modern India. Only some relevant documents have been preserved due to personal efforts. For example, private papers of M.N. Saha were preserved by his children after his untimely death. The original papers of Saha were deposited in the Nehru Memorial Museum and Library, with a complete set of copies of these papers kept at the Saha Institute of Nuclear Physics in Calcutta. Based on the Saha papers, I gave an account of the intellectual growth of Saha, describing how he formulated the ionization equation [24]. However, it is difficult to carry on similar studies for other Indian physicists of that era due to the lack of appropriate source materials. C.V. Raman also had the habit of keeping his personal papers organized systematically.  I have talked with persons who had seen the huge collection of Raman's private papers at the Raman Research Institute even a few years after his death.  But then these papers disappeared mysteriously! Scholars who have studied Raman's scientific works debate as to what happened to his private papers – with no answer available so far.  It is a national shame that we have not saved the documents which may be needed to write a complete history of the only Nobel-winning scientific work done so far in India. It should also be emphasized that merely preserving the source materials is not enough. They have to be properly catalogued and made available to scholars who want to use these for their research.  It is believed that J.C. Bose's private papers are preserved.  But generations of scholars who wanted to consult these papers for historical research have been denied access to them. One ray of hope is a very recent initiative at the Birla Industrial and Technological Museum, Kolkata, to set up an archive for scientists who worked in eastern India.

In the era of European colonial empires, no other colony achieved anything in basic sciences comparable to what India achieved.  Much of the source materials which might have been essential to reconstruct a proper history of this extraordinary era of Indian science have been lost.  There are still not enough serious efforts to preserve what still remain, and what still remain are likely to be lost in the coming years, unless a national initiative is taken to preserve these urgently – and very soon.

**Acknowledgements.**  I thank Sekhar Bandyopadhyay and Tulasi Parashar for urging me to give a seminar at the New Zealand India Research Institute at Victoria University of Wellington explaining the physics breakthroughs in colonial India to non-experts.  Afterwards, I have given this seminar, on which this paper is based, in several other places. I thank Robert Anderson, Syamal Chakrabarti, Enakshi Chatterjee, Deepanwita Dasgupta, Subrata Dasgupta, Deepak Kumar, D.C.V. Mallik, Suprakash Roy and Rajinder Singh for many discussions on the history of science in colonial India over the years. I am grateful to K. Indulekha for going through the manuscript very carefully and for suggesting improvements.